\documentstyle[12pt,psfig]{spacekap}
\def\farcs{\hbox{$.\!\!^{\prime\prime}$}}  
\def\farcm{\hbox{$.\!\!^{\prime}$}}
\def\sss{\hbox{$.\!\!^{s}$}}

\def\asec{\ifmmode ^{\prime\prime}\else$^{\prime\prime}$\fi}
\def\lea{\ifmmode ^{<}_{\sim} \else $^{^{<}_{\sim}}$\fi}
\def\gea{\ifmmode ^{>}_{\sim} \else $^{^{>}_{\sim}}$\fi}
\begin{opening}
\title{
                       Optical observations                       
                of nearby isolated pulsar PSR 0656+14                
                     at the 6-meter telescope.                    
}
\author{V.G.Kurt}
\author{B.V.Komberg}
\institute{ Astro Space Center of Russian Acad. of  Sci.,  117810,  Moscow,
Russia }
\author{V.V.Sokolov}
\author{S.V.Zharykov}
\institute{ Special Astrophysical Observatory of Russian Acad. of Sci,
Karachai Cherkessia, Nizhnij Arkhyz, 357147,Russia;
E-mail: sokolov@sao.ru}
\date{}
\end{opening}
\runningtitle{Optical observations of the pulsar PSR 0656+14}
\runningauthor{V.G.Kurt, B.V.Komberg, V.V.Sokolov, S.V.Zharykov}
\begin{document}
\begin{abstract}
The data of {\it BVR} observations of the middle-age radio pulsar PSR 0656+14
on January, 20/21 at the BTA (6-m) are presented. 
The brightness is determined in Cousins {\it B} filter $B\approx 25.1$
with $\lambda_{eff}=4448 \rm \AA$ in adjacent for HST F130LP long-pass filter of a
star-like object, coinciding with the position of VLA radio source.
Relatively large observed {\it V} and {\it R} fluxes
($\lea 3\sigma\ or\ > 10^{-30}\ ergs\ cm^{-2}\ s^{-2}\ Hz^{-1}$)
can witness a non-thermal
nature of optical radiation of this pulsar
up to $\lambda \approx 6600\ \rm \AA$.
Most probably, in the UV-optical ({\it BVR}) spectral range a power-law spectrum
is superimposed on the thermal-like radiation of the entire
neutron star surface what can be related to a mechanism itself of
the pulsar activity.

\end{abstract}

\section{ Introduction}

     At present above 700 radio pulsars are  discovered,  but  the
optical radiation can be assumed reliably detected only  for  some 
of them: the famous Crab pulsar, PSR  0540-69,  PSR  1509-58,  the 
Vela pulsar, the gamma-ray + X-ray pulsar  Geminga,  and  for  PSR 
0656+14 (\cite{12}; \cite{4};
 \cite{13}; \cite{9};  \cite{3};
 \cite{5};  \cite{11}).
  I.e.  the  probable   optical
companions are already detected  the  which behavior can  be 
investigated by more careful study of their spectra  and  temporal 
variability  both  in  X-ray,  gamma,  radio,  and  optical.  Recent 
observations at the Hubble Space  Telescope  (HST)  in  UV-optical 
range led to the detection of  corresponding  probable  UV-optical 
counterparts  of another two isolated pulsars PSR 0950+08 and  PSR 
1929+10 (\cite{11}).

     PSR 0656+14 was also identified recently in  optical  (\cite{4})
 and  immediately  after  that  the  study  of  its
spectrum was started at the HST (\cite{11}).  At  the
6-meter telescope (BTA) this pulsar and others is studied  within  the 
framework of a program of  wide-band  ground-based  photometry  of 
nearest pulsars of the Northern sky.  For  a  middle-aged  pulsar
PSR 0656+14 ($\tau =\frac{P}{2\dot{P}} = 110000$ yr) we  suppose  
to  fulfill  a
multi-color photometry for the purpose of refinement of the nature 
of optical radiation  of  the  isolated  neutron  star  (INS),  by 
supplementing  HST  UV-optical  observations   with   ground-based 
{\it BVRI}-observations.

     Certainly, the basic (global) goal pursued by many groups  at 
the investigation of INSs-pulsars is to select a thermal component 
of radiation (including the optical one) for the  pulsars  of  the 
age of $>~10^5$ years arising from the entire neutron star  surface.
This problem is still actual since  as  is  noted  in  all  recent 
papers on INSs it would allow us (together with the study in {\it EUV}
and X-ray ranges) to refine the thermal evolution of these compact 
objects and to approach in the end the correct equation  of  state 
of matter for their interior regions with supernuclear densities.
 A possibility is now actively discussed of a presence in  the
deep interior of "neutron" stars of a pion or quark condensate , a 
superfluidity, and others.  See,  for  example,  Umeda, 
Tsuruta, and Nomoto (1994); Meyer, Pavlov,  and  Meszaros  (1994). 
Though, it should be remarked that UV-optical   thermal  radiation 
was apparently observed only for two active and more old pulsars:  
PSR  1929+10 and (probably) PSR 0950+08 (\cite{11}).
In other cases the thermal radiation from INSs is basically observed 
in {\it EUV} and soft X-ray bands.

     The  study  of  spectra  of  the  nearest  isolated  pulsars, 
including the X-ray brightest PSR  0656+14,  was  begun  with  the 
study in X-ray (\cite{14}). In particular, now for PSR
0656+14 the observation in soft X-ray  range  of  the 
ROSAT observatory (\cite{7}) are  also  used.  The
high quality of X-ray spectra allows to determine rather precisely 
an effective temperature of the radiating  surface  of  a  neutron 
star by means of "fitting" the observed spectra to fit  the  black-body
like  radiation  in  Wien  region.  However,  the  most   recent
observations of isolated pulsars, such as Geminga and PSR 0656+14, 
including optical investigation (\cite{3};  \cite{11})
 showed that  in  optics  the  effects  can  become
essential  which  are  related   either   to   the   presence   of 
geometrically thin ($\sim 1.5$ cm), but optically thick  atmosphere,  or
with the influence of magnetic field ,  or  with  the  non-thermal  
contribution of radiation from polar  caps,  or  with  some  other 
non-thermal effects. One way or  another,  it  turns  out  that  a 
simple black-body fitting an  observed  spectrum  can  be  quite 
non-adequate to not only X-ray range (\cite{10}),  but
especially to optical range (\cite{3};  \cite{11}).
 I.e. in the UV-optical range a non-thermal  radiation
could dominate the optical spectrum at  least  in  the  middle-aged
pulsars. 

     Though the study of non-thermal radiation is  interesting  by 
itself  from  the  point  of  view  of  elucidation  of   physical 
conditions in pulsar magnetospheres and refinement of a theory  of 
pulsar emission,  but  it  is  nevertheless  a  "barrier"  in  the 
movement to the basic goal -  the  elucidation  of  the  main question:
"What the interior of neutron stars  consist  of ?"  On  the  other
side, hopefully, the presence of a just non-thermal  component  of 
radiation  can  increase  considerably  the  luminosity  of  these 
objects in optical.  The  last  was  also  directly  confirmed,  in 
particular, by our BTA observations of PSR 0656+14 in {\it B, V, R} filters,
about what the 2-d section of this paper says. We obtained for the first 
time the estimations of the brightness in this bands from the Earth. 
Though in {\it V} filter the observational material of approximately the
same quality  was  already  obtained  at  two  ESO  telescopes  by 
\cite{4}.

     The first attempt  to  observe  PSR  0656+14  in  optical  was 
undertaken by Cordova {\it et al.} (1989) after the identification  of
this pulsar in X-rays. Though an optical counterpart was not  then 
detected, this observation showed that  the  corresponding  region 
around the VLA position of the pulsar is not espacially crowded and 
does not contain too bright objects nearer $\sim 5$ arcsec from the pulsar.
Thereupon   a   successful   ground-based    optical 
observation of PSR 0656+14 which is the X-ray brightest  from  all 
"normal" radio pulsars was fulfilled in  1989  at  the  3.6-m  ESO 
telescope for a total exposure time of 60 minutes in {\it V} filter  and
in 1991 with the NTT for  the  70  minutes  total  exposure  in {\it  V}
bandpass (\cite{4}).  In  both  cases  the  authors
detected an object which coincides well with VLA position  of  PSR 
0656+14  as  measured  by  Thompson  and   Cordova   (1994).   The 
corresponding object has $V \sim 25$ with the error of  0.5  mag,  what
corresponds to the 3$\sigma$ level of detection.  However,  a  large
stellar   magnitude   of   optical   counterpart   together   with 
uncertainties in the estimate of distance to the  pulsar  (100-700 
pc) makes difficult the classification of optical both thermal and 
non-thermal  radiation,   especially   in   terms   of   uncertain 
interpretation of X-ray data (\cite{7}).

     New observations of this pulsar were carried out at  the  HST 
with the UV-sensitive Faint Object  Camera  by  Pavlov  {\it et  al.,}
(1996). The observations of the pulsar candidate for  PSR  0656+14 
were fulfilled in F130LP filter with the band width  $\lambda\lambda
= 2310 -4530 \rm \AA$, with center  at  the  $3365 \rm \AA$,  including  the
radiation in standard {\it U} and {\it B} filters. The exposure was 4755  sec.
Near the VLA position of PSR 0656+14  in  the  deep  $7\farcs4\times7\farcs4$
images there is only one point-like object with $m_{130LP} = 25.19\pm
0.04$ ($S/N=52$). Results of first observations at the 6-m  telescope
in more narrow spectral bands for the candidate identification for 
PSR 0656+14, supplementing the observations of the group of Pavlov 
{\it et al.} (1996) in space are presented in Section 2.

     Thus, the goals of this paper are:
1). to confirm by our observations a non-thermal nature of  optical 
radiation of the candidate identification for PSR 0656+14;  
2). to understand also how the optical spectrum of this middle-age 
pulsar differs from an analogous spectrum of Geminga (\cite{3})
 which is much alike to it - another middle-age INS. To
draw a power low spectrum Pavlov {\it et al.} (1996) used the point of
$V\approx 25$ obtained in observations at NTT by Caraveo  {\it  et.al.} 1994)
and their own point obtained at HST in F130LP filter. On  the  one 
hand, the measurements of  brightness  of  the  optical  candidate 
identification for PSR 0656+14, which has been gained in "grond-based" {\it B} band,
confirmed indeed a non-thermal nature of optical radiation of this 
pulsar in optical. On the other hand, {\it BVR} observations at  the  6-m
telescope give an opportunity to say more reliably about a power-law 
spectrum in optical, but not  about  a  cyclotron  feature  on  the 
thermal continuum, which is the Geminga case. Section 3  describes 
that in more details. 

     The conclusion notes that in all cases of observations at the 
4 telescopes and in  different  optical  spectral  bands  the  PSR 
0656+14 turns out to be 1.5-2 stellar magnitude brighter than  was 
expected before that on the basic of simple black-body fitting of 
X-ray and optical data.

\section{ The Observations.}

\begin{figure}[t]
\centering{
\vbox{\psfig{figure=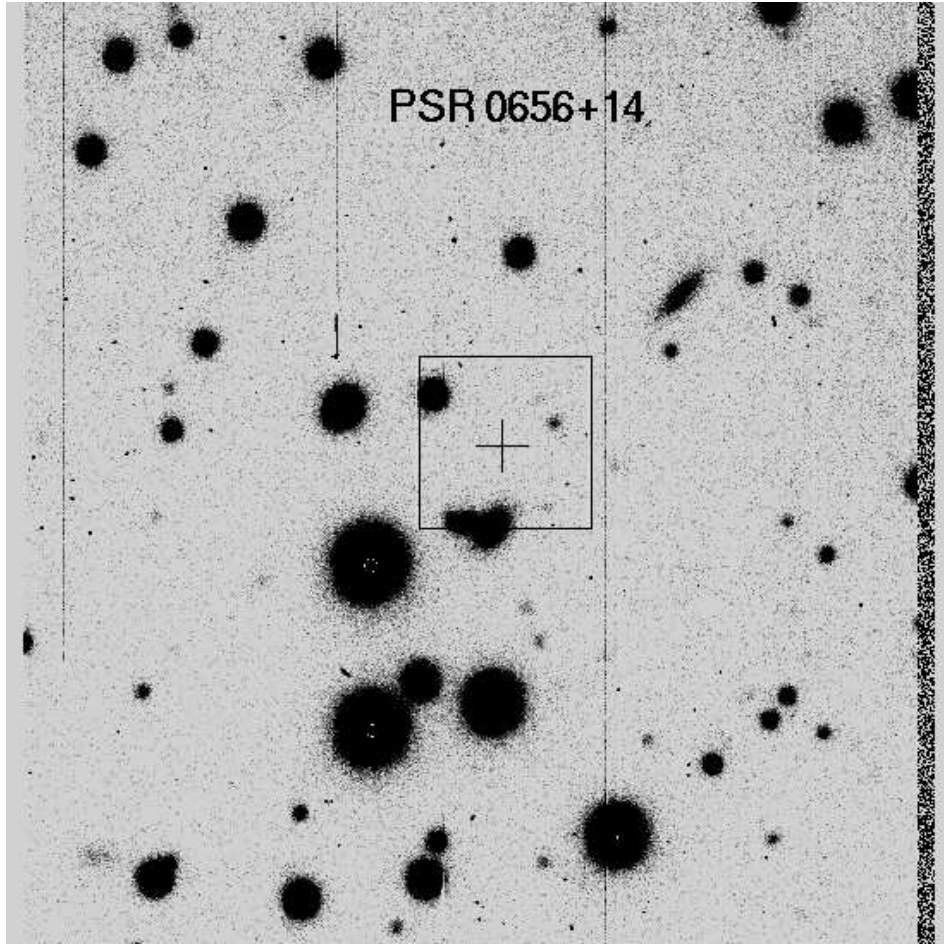,width=12.5cm,%
 bbllx=70pt,bblly=300pt,bburx=520pt,bbury=755pt,clip=}}  \par
}
\caption{
 The {\it B} field of PSR 0656+14 with the image size of $2\farcm38\times2\farcm66$.
The North is top, the East is left. The cross denotes the VLA
position of the pulsar. The square denotes a fragment shown in
Fig.2.
}
\label{psr1}
\end{figure}

   The photometric observations of PSR 0656+14 for the purpose  of 
detecting and estimating the  brightness  in  {\it B, V}  and {\it R}  Cousins
filters for PSR 0656+14 were carried out at the 6-meter  telescope 
of SAO RAS on January, 20/21, 1996 with CCD  photometer  installed 
in the primary focus. We used CCD "Electron ISD017A"  of  size  of 
$1040\times1160$ pixels, what corresponds to the area of $2.38\times2.66$
arc min in the sky sphere. The CCD photometer was used in  such  a 
mode  ("bining  $2\times2$"),  that  an  effective  CCD   size   was
$520\times580$ elements with the size of a separate element  being 
$0.274\times0.274$ arc sec. 

   We obtained 4 exposures each 600 s in  {\it B}  filter,  4  exposures
each 600 s in {\it V} filter and 4 exposure each  400  s  in  {\it R}  filter.
Unfortunately, durin approximately two hour  of  observations  the 
atmospheric transparency was changing considerably, and the seeings  
were $\approx  2\asec$. Nevertheless, in the place of the best VLA position  of
PSR 0656+14 , where a $V\approx 25$ object was  already  detected  (\cite{4}),
 approximately at  the  same  level  $\approx  3
\sigma$ an object is also detectable on the sum of  all  exposures 
in each filter. In {\it B} filter on the  sum  of  all  exposures a weak
object is obviously present at the level of $\geq 3\sigma$, the
brighness estimate of which is gained to  be  fulfilled  with  the 
precision not worse than it was done at NTT by Caraveo  {\it et. al.,} (1994)
for {\it V} band.

   The standard processing of data included the subtraction of  so 
called electronic shift - and additive  component  of  CCD  result 
signal -  and  the  division  into  "the  flat  field",  i.e.  the 
correction of non-uniform sensitivity of  detector  elements.  The 
traces of  space  particles  were  eliminated  by  the  method  of 
interpolation between neighbour values. Since  in  the  moment  of 
observation the atmospheric conditions were not  photometric,  the 
photometric  calibration  of  the  object  in  the  area  of   the
localization of PSR 0656+14 was carried out with the help of   CCD 
photometric observation (in the same {\it BVR} Cousins  system)  at  the
1-meter Zeiss-1000 telescope of the SAO RAS in March 1996 in a good
photometric night. Astrometric referencing of the pulsar  position 
on the obtained CCD images was fulfilled by Digital Sky Survey
(DSS), and also by optical data published  by  Caraveo  {\it et  al..}
(1994). All processing was made with the  use of a software 
"MIDAS". 

\begin{figure}[t]
\hspace{2cm}
\centering{
\vbox{\psfig{figure=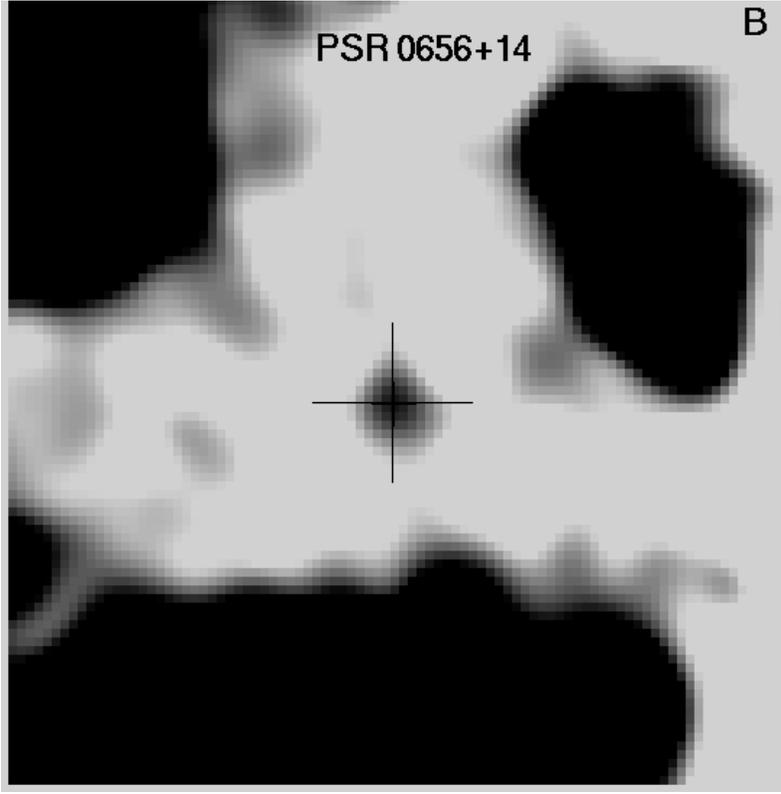,width=10.5cm,%
 bbllx=70pt,bblly=300pt,bburx=520pt,bbury=755pt,clip=}}  \par
}
\caption{
 The fragment $25\asec\times25\asec$ of the sum {\it B} image with the candidate
identification for PSR 0656+14 in the center. The exposure time
in {\it B} filter was 2400 s for the seeing of $2\farcs1$. A result of Gauss
smoothing is shown. The source coinciding with the VLA position
has $\approx25.1$.
}
\label{psrB}
\end{figure}

   Figure 1 shows the image - one of 600th second  exposure  in {\it B}
filter. The field of PSR 0656+14 with the frame size  of  $2\farcm38\times
2\farcm66$. N to the top, E to the left. The cross corresponds  to  the
position of {\it V}-candidate identification for PSR 0656+14  using  the
two images obtained by by Caraveo {\it et al.}.(1994). This  position  is
determined by bright stars  with  an  accuracy  of  a  pixel  size 
($0\farcs274$) in our image. The square marks  a  fragment  of  the  PSR
0656+14 field, the sum image of which is shown in Fig.2. The  size 
of the  fragment  is  $25\asec\times25\asec$.  Fig.2  shows  a  result  of  Gauss
smoothing by a sum of four exposures in {\it B} filter  with  the  total
exposure of 2400 s.  A  faint  source  with  $B  =  25.1\pm0.5$,
coinsiding with the VLA position of  PSR  0656+14:  $\alpha(1950)  =
06^{h} 56^{m} 57\sss942$; $\delta(1950) = 14^o 18^{\prime} 33\farcs8$ is in the  center
of the fragment. Although detected only at $\approx 3\sigma$  level,  its
reality is not in doubt. It should be  noted  that  the  effective 
time of observation of the  object  of  such  a  magnitude  and  a 
corresponding value of Signal/Noise ratio for a good  trancparency
and at seeings not worse than $1\farcs5$ is less than 10 min.

   Thus, because of the bad weather conditions the photometric estimate 
of brightness corresponding to extra-atmospheric 
Log Flux ($erg\  cm^{-2}\  s^{-1}\  Hz^{-1}) = - (29.43\pm0.20$) which we use in
Section 3 for "the joint" of ground-based {\it BVR} observations
and HST F130LP observations in frequency adjacent F130LP and our
Cousinse B filter turns out to be the most reliable only for {\it B}
filter. ( For the Cousins {\it B} filter: $\lambda_{eff} = 4448 \rm \AA$
and $FWHM = 1008 \rm \AA$.) Since in the sum {\it V}-images and {\it R}-images
"hot pixels" are observed at the level of $\lea3\sigma$ at the same place,
now we can only say that most probably the observed object does
not obey the Rayleigh-Jeans-like law in these bands also. Here the
fluxes are obviously greater than the $10^{-30}\ erg\ cm^{-2}\ s^{-1}\ Hz^{-1}$.

   We fulfilled the  pass  from  the  stellar  magnitudes  of  the 
Cousins system to the absolute fluxes in $erg\ cm^{-2}\ s^{-1}\ Hz^{-1}$ with
the use of data on $\alpha$Lyr  published  by  Fukugita  {\it et.al.}
(1995).

\section{Discussion.}

\begin{figure}[t]
\hspace{2cm}
\centering{
\vbox{\psfig{figure=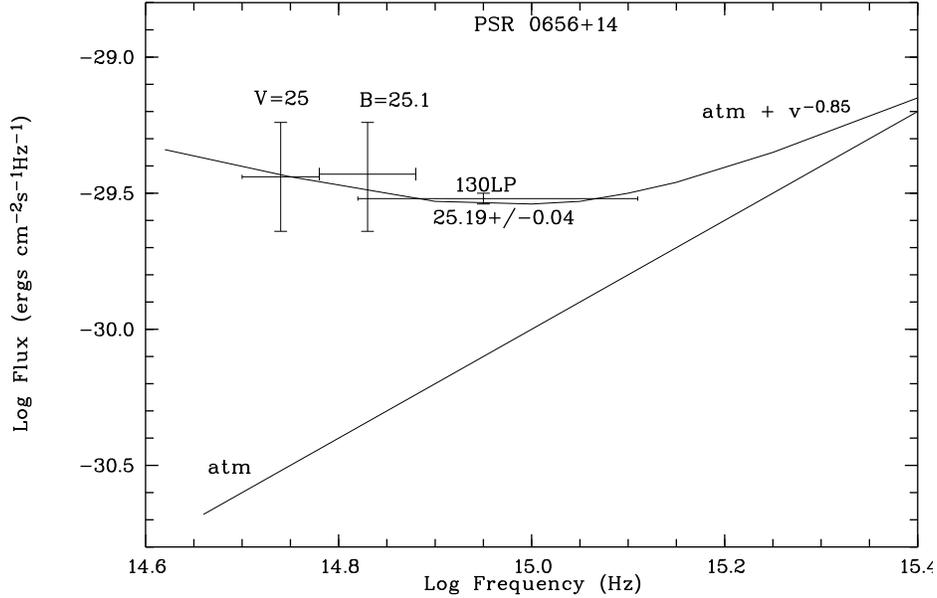,width=12.5cm,%
 bbllx=0pt,bblly=40pt,bburx=580pt,bbury=430pt,clip=}}  \par
}
\caption{
    Its shows the  "{\it B}-joint"  of  extra-atmospheric  (Pavlov, G.G. {\it et.al.} 1996)
  and  the  ground-based  photometric  data  known  by
September 1996. The wide cross shows the energy flux detected with
the HST F130LP ({\it B+U}+...) filter, which, beside HST  UV  radiation,
transmits also the BTA(6m) "ground based" {\it B}  optical  quanta.  The
cross labeled "$B=25.1$" shows the result of a photometric  estimate
of the brightness of PSR 0656+14 candidate in the Cousins {\it B} filter
at the 6-m (BTA) telescope of January, 25/26, 1996. The cross with
"$V=25$" shows the NTT result by Caraveo {\it et al.} (1994). The
line labeled "$\nu^{-0.85}+atm$" shows a fit the UV-optical (HST +
ground-based) data for these 3 spectral points.The straight line labeled  
"atm" shows separately the magnetic model atmosphere thermal-like fluxes 
for effective temperature $T=530000$ K at the entire neutron star 
surface and for $d=280$ pc.  
}
\label{psrB}
\end{figure}

   In Fig.3 all extra-atmospheric UV-optical fluxes  detected  for 
the  PSR  0656+14  candidate  in  different  badpasses  known  by
September 1996 are shown. The wide cross  shows  the  energy  flux 
detected with the F130LP filter of the UV-sensitive  Faint  Object 
Camera on-board HST. The cross labeled "$B=25.1$" shows  the  result
of photometric estimate of the PSR 0656+14 candidate brightness in 
the Cousins {\it B} filter at the 6-m telescope on January, 25/26, 1996.
The cross with "$V=25$"  shows  the  results  by  Caraveo {\it et  al.}
(1994). The  line labeled "$\nu^{-0.85}+atm$" shows a fit the
UV-optical (space  +  ground-based)  data  for  these  3  spectral 
points.

   Here we give one of possible interpretations of thermal soft X- 
ray ROSAT and UV-optical  radiation  for  PSR  0656+14  candidate, 
supposed in the paper by Pavlov {\it et al.} (1996). The  curve  
"$\nu^{-0.85}+atm$" is a two-component model, combining a power-law  with  a
magnetic model atmosphere spectrum. The spectrum corresponding  in 
the UV-optical range to only this  magnetic  model  atmosphere  is 
shown separately. It is  the  straight  line  labeled  "atm".  This 
radiation corresponds to the thermal-like flux, arising from  the 
entire neutron  star  surface  with  atmospheric  model  effective 
temperature  (observed  from  the  infinity)  $T=530000$ K  and   for
$d=280$ pc.  This  surface  thermal-like  radiation   seems   to   be
observable only in {\it EUV} and in the soft X-ray ranges and the  "atm"
line is obtained by fitting  the  thermal  soft  X-ray  ROSAT  PSR 
0656+14 candidate spectrum by different magnetic atmosphere models 
for a given distance {\it d} to the source.

   The magnetic field is $B=4.7\cdot10^{12}$G, $M=1.4M_{\odot}$,
("true" radius)/(the radius for infinity) = 10km/13km.  These  are 
the parameters of a model from the paper by Pavlov {\it et al.,} (1996;
and references therein) used for the interpretation of observed spectrum 
of the PSR 0656+14 candidate at $d = 280(+60;-50)$pc (\cite{1}).
Though, generally speaking, a very soft X-ray radiation in correspondence 
with small interstellar absorption  allow  a  supposition  of  the 
distance to the X-source E0656+14 to be somewhat between  100  and 
600 pc (\cite{7}). And the estimate  of  distance  to
PSR 0656+14 by dispersion measure gives{\it d} up to 760 pc (\cite{15}).
 But nevertheless, 600-700 pc seem  to  be  too  large
distances, since both a  measured  proper  motion  of  the  pulsar 
(\cite{15}) and a very low X-ray absorption  allow
a possibility of approximately the same {\it d},  as  in  the  case  of
Geminga object, for which the direct measurements of parallax give 
now the distance of $\approx 160$ pc (\cite{3}).

   As is seen from Fig.3, our estimate of brightness  of  the  PSR 
0656+14 optical candidate in  more  narrow  "ground-based"  {\it B}-band
confirms once more a non-thermal character of UV-optical  spectrum 
of this object even with such uncertainties in {\it B} flux. It could be
supposed that the contribution of neighbor objects in Fig.2 in the 
vicinity of our PSR 0656+14 optical  candidate  does  not  distort 
this estimate, since  "the  deviation"  from  Rayleigh-Jeans-like 
atmosphere model spectrum begins already in the  wide  HST  F130LP 
filter (transmitting also BTA {\it B} "ground-based"  quanta)  for  the
only bright ($S/N=52$) point-like object in the Plate 1(a) from  the
paper of Pavlov {\it et al.,} (1996). The last  means  that  our  data
"joint" ground-based and HST observations, what in total allows to 
say reliably that in all the  three  cases  the  same  object  was 
observed. 

   Thus, an optical object coinciding with VLA-position turned out 
to be brighter indeed than was  expected  for  purely  black-body 
dependence Flux/Frequency in optical by BTA observations  also.  At 
the conditions under which the pulsar was observed in January 1996 
it could not be detected at all if it is more than 2 st.mag. fainter 
indeed, as follows from Fig.3. January 1996  observations  carried 
out at such weather  conditions  were  meaningful  either  if  the 
optical spectrum is non-thermal, or if the object is  much  nearer 
(than 600 pc) if we still deal with the black-body like  radiation
of the surface.  

   It should be said that our original goal was indeed  to  detect 
the thermal-like radiation from the  neutron  star  surface.  This 
non-thermal "interference", like the case of Geminga (\cite{3}),
  remains  a poor interpreted   surprise   (\cite{11}).
 Though the distance to the  source  seems  to  be
less than 500 pc indeed, nevertheless the direct  measurements  of 
parallax are necessary since  the  exact  distance  increases  the 
reliability of the temperature estimate of the entire neutron star 
surface, made by Pavlov's group. The exact distance to the  object 
allows also a better correction of the spectrum, as it can be done 
now for Geminga. 

   It should be also necessarily said that for the Geminga object, 
which is much alike to the PSR 0656+14, the  situation  in  optics 
looks somewhat different. In  our  case  the  spectrum  obeys  the 
Pavlov's power spectrum in Fig.3 and  (together  with  HST  F130LP 
observations) it does not show a sharp deficit of flux  in  the  {\it B}
band in comparison to the {\it V} brightness which apparently is observed
for Geminga (\cite{3}). (It should be noted that  the
statement or suggestion about the cyclotron feature for 
Geminga is based on the {\it B} and {\it I}-ground-based data of Bignami {\it et.  al.}
(1988,1996)  with  the  $B=26.5\pm0.5$.).  From  our  January   1996
observations of PSR 0656+14 in {\it V} and {\it R} bands it follows that the {\it V}
brightness is close indeed to what Caraveo {\it et  al.}  (1994)  give
for the PSR 0656+14 optical candidate. The object is seen  at  the 
6-m telescope in {\it R} filter with the  flux  obviously  greater  than
$10^{-30}\ erg\ cm^{-2}\ s^{-1}\ Hz^{-1}$. The total means that this flat  power
spectrum of the PSR 0656+14 candidate begins in the  HST  UV-range 
and can be continued at least up to {\it R}-optical range.

\section{Conclusions}

   So, from the result of observation at 4-th (3.6-m ESO,  NTT,  HST, 
BTA) telescopes  in  different  optical  spectral  bands  the  PSR 
0656+14 candidate turns out to be 1.5-2 stellar magnitude brighter 
than was expected earlier, proceeding from  a  simple  black-body 
fit of X-ray + optical data. The radiation of  PSR  0656+14 
is basically non-thermal  indeed  in  optical,  though  the  entire 
neutron star surface radiation can become dominating in the far UV 
range. As to our {\it B}  estimate  of  brightness,  together  with  HST
F130LP and ground-based observations by Caraveo {\it et  al.,}(1994) in  {\it V}
filter, the non-thermal spectrum can be  approximated  by  one  of 
power lows suggested in the paper by Pavlov {\it et al.,} (1996). Most
probably this object has no sharp decrease of  flux  in  the  more 
narrow (in comparison to F130LP) "ground based" {\it B} band as compared
to the flux in {\it V}  and {\it  R}  filters,  which (a  dip) is apparently
observed from Geminga (\cite{3}).

   To confirm or to rule out the tendency of  continuation of  the 
PSR 0656+14 optical candidate spectrum into the red range with one 
or another power-law and also to have reliable data for development
of quantitative models of non-thermal radiation from  INSs  it  is 
necessary to  carry  out  additional  {\it BVRI}  observations.  Further
multicolor photometry would be needed. In  particular,  the  Crab- 
pulsar spectrum is studied rather in details in this range and  it 
is flat indeed with $\alpha =  -0.11\pm0.13$  (\cite{12}),
though all other features of  these  two  pulsars are  much
different. 

   The authors sincerely thank George Pavlov for  the  most  fresh 
information on the HST observations of INSs and for sending us the 
paper of Pavlov {\it et al.,} (1996) prior to  publication.  We  thank
also Alexander Kopylov for the help  in  observations  and  active 
discussions of the obtained results and Tatyana Sokolova for  
preparing this text to publication.

\end{document}